\documentclass[12pt,preprint]{aastex}

\begin{document}

\title {A Study of Gravitational Lens Chromaticity using Ground-based Narrow band Photometry}


\author{A. M. Mosquera$^{1,2}$, J. A. Mu\~noz$^1$, E. Mediavilla$^{3,4}$, and C. S. Kochanek$^{2,5}$}

\bigskip

\affil{$^{1}$Departamento de Astronom\'{\i}a y Astrof\'{\i}sica, Universidad
       de Valencia, E-46100 Burjassot, Valencia, Spain}
\affil{$^{2}$Department of Astronomy, The Ohio State University, 140 West 18th Avenue, Columbus, 
OH 43210, USA}
\affil{$^{3}$Instituto de Astrof\'{\i}sica de Canarias, E-38200 La Laguna,
       Santa Cruz de Tenerife, Spain}
\affil{$^{4}$Departamento de Astrof\'{\i}sica, Universidad de La Laguna, E-38200 La Laguna, 
       Santa Cruz de Tenerife, Spain}
\affil{$^{5}$ Center for Cosmology and Astroparticle Physics, The Ohio State University,
191 West Woodruff Avenue, Columbus, OH 43210, USA}


\begin{abstract}

We present observations of wavelength-dependent flux ratios for 
four gravitational lens systems (SDSS~J1650+4251, HE~0435$-$1223, FBQ 0951+2635, and Q~0142$-$100) 
obtained with the Nordic Optical Telescope. The use of narrowband photometry, as well as the excellent seeing conditions during the 
observations, allows us to study their chromatic behavior. 
For SDSS~J1650+4251, we determine the extinction curve of the dust in the $z_L=0.58$ lens galaxy 
and find that the 2175 \AA \ feature is absent. In the case of  HE~0435$-$1223, we clearly 
detect chromatic microlensing. This allows us to estimate the wavelength-dependent size 
of the accretion disk. We find an R-band disk size of $r^{R}_s=13\pm5$ light days for a 
linear prior on $r^{R}_s$ and $r^{R}_s=7\pm6$ light days for a logarithmic prior. 
For a power-law size-wavelength scaling of $r_s\propto\lambda^{p}$, we were able to constrain the 
value of the exponent to $p=1.3\pm0.3$ for both $r^{R}_s$ priors, which is in agreement with 
the temperature profiles of simple thin disk models ($p=4/3$).

\end{abstract}


\section{Introduction}

Gravitational lenses can act in many cases as cosmic telescopes. Since they can 
magnify the flux of background sources, they allow us to look far away in 
cosmic time, and to study in more detail the physics of the universe at cosmological 
distances, and beyond the resolution capabilities and detection thresholds of 
current telescopes. In particular, in gravitationally lensed quasars where variations in color 
are observed between their images, the ``chromaticity'' can be used to 
probe active galactic nucleus (AGN) and galactic structure using microlensing 
(see the review by Wambsganss 2006) and to study the properties of 
dust in intermediate- and high-redshift galaxies, as these are the effects leading 
to color differences between lensed images (see the review by Kochanek 2006).  

The extinction effects are due to differences in the amounts and properties of 
dust in the lens galaxy near each image (Nadeau et al. 1991). In many cases the 
extinction curve of the dust in the lens galaxy can be accurately determined by measuring the 
wavelength dependence of the image brightnesses. This method is 
similar to the ``pair method'' (Massa et al. 1983), comparing 
the spectral energy distributions of stars with the same spectral type but different reddenings, 
used to study dust properties in the Milky Way (MW) 
and in nearby galaxies where the stars can be individually resolved. In the MW, most of 
the observed extinction curves are well fit by the so-called Cardelli et al. (1989, hereafter CCM) 
parameterization, a relationship that only depends upon the ratio 
of the total to the selective extinction, 
$R_V$. This parameter can take values from $2.1$ to $5.8$ depending on 
the line of sight, with a typical value of $R_V\sim 3.1$ (Draine 2003). 
The CCM parameterization includes a spectral feature at 2175 \AA, but with 
some important differences that depend on environment. The Large Magellanic Cloud, for example, 
shows a weaker 2175 \AA~ feature in some regions (e.g., Nandy et al. 1981; Misselt et al. 1999; 
Gordon et al. 2003), while it is entirely absent for dust in the bar of the Small Magellanic 
Cloud (SMC, e.g., Pr\'evot et al. 1984; Gordon et al. 2003). The (classical) ``pair method'' 
has been applied successfully up to distances $\sim 10$ Mpc (e.g., Bianchi et al. 1996), but cannot be 
used at longer distances because it requires observations of individual stars.
 
At cosmological distances, the pairs of lensed quasar images provide a powerful 
extension of the classical ``pair method''. Falco et al. (1999) studied extinction 
properties for a sample of 23 intermediate-redshift galaxies. They were able to estimate 37 
differential extinctions, and also some absolute extinctions when the total absorption was 
large enough. The first measurement of an extinction curve at high redshift of 
comparable quality to those obtained in the MW was obtained by Motta et al. 
(2002). Mu\~noz et al. (2004) estimated the extinction laws in two intermediate-redshift 
galaxies and found unusual extinction curves compared with the MW. Mediavilla et al. (2005) 
extended the study of dust extinction properties to the far-UV, where extinction is strongest 
and different behaviors have been observed in different environments (Gordon et al. 2003). 
El\'{i}asd\'ottir et al. (2006) used theoretical analysis and simulations to study the 
effects of extinction in 10 intermediate-redshift galaxies and found no evolution
of the dust properties with redshift even though, as they point out, a larger sample of 
lenses would be needed to reach a robust conclusion. Dai $\&$ Kochanek (2009) measured 
the dust-to-gas ratio in six distant lens galaxies using {\it Chandra} observations to 
measure the difference in the gas column for comparison to the extinction differences, and 
they found a value consistent with the average ratio in the MW.

There are few other methods for studying dust extinction 
outside the MW and nearby galaxies. Another ``non-classical'' 
pair method compares reddened and unreddened photometrically similar 
Type Ia supernovae(SNe Ia; Riess et al. 1996; Perlmutter et al. 1997). 
Quasars with foreground absorbers have been used to 
measure the excess reddening produced by these systems. For instance, Murphy $\&$ Liske 
(2004) studied the properties of dust associated with damped Ly$\alpha$ systems. 
Galactic environments were also traced with absorbers such 
as CIV (Richards et al. 2001), CaII (e.g., Wild et al. 2006), and MgII 
(e.g., M\'enard et al. 2008). Finally, the spectra of the afterglows of gamma-ray 
bursts (GRBs) have also been used to determine extinction curves of dust in the 
host galaxies (e.g., Jakobsson et al. 2004). These methods have found 
that extinction curves similar to those in our neighbor galaxies are also found 
at high redshift. The 2175 \AA~ feature has been confirmed in several systems (Motta et al.
2002; Junkkarinen et al. 2004; Wang et al. 2004; Ellison et al. 2006), and 
its highest redshift ($z=2.45$) detection to date was made by El\'{i}asd\'ottir et al. (2009) 
using GRBs. The SMC extinction law also appears to reproduce the extinction in some intermediate- 
and high-redshift systems, such as the line of sights of many quasars (e.g., Hopkins et al. 2004), 
and to other GRB afterglows (e.g., Kann et al. 2006), although the sample of robustly 
confirmed SMC-like extinction laws is very small. 

Besides extinction, microlensing-induced variability by stars and compact objects in 
the lens galaxy (see the review by Wambsganss 2006 and references therein) can 
also produce chromatic variations between the quasar images. The magnification 
produced by microlensing depends strongly upon the source size, with smaller 
emission regions showing larger variability amplitudes. In particular, since 
quasar accretion disk models (Blaes 2004, and references therein) all assume that 
the thermal emitting region is larger at longer wavelengths, we should observe 
different microlensing magnifications at different wavelengths (i.e., chromatic microlensing). 
Although non-negligible, this chromatic variation 
is more difficult to observe compared with achromatic microlensing, because microlensing 
magnification gradients on the scale of the source size are required to produce the 
effect (Wambsganss $\&$ Paczynski 1991). However microlensing-induced chromaticity has been 
detected and analyzed in many lens systems (Anguita et al. 2008; Bate et al. 2008; 
Eigenbrod et al. 2008; Poindexter et al. 2008; Floyd et al. 2009; Mosquera et al. 2009; 
Blackburne et al. 2011). The sizes of the non-thermal emission regions have also been studied 
using the differences between optical and X-ray flux ratios (Pooley 
et al. 2007; Morgan et al. 2008; Chartas et al. 2009; Dai et al. 2009; Blackburne et al. 2011). 
These results showed that microlensing is a unique tool to zoom in on AGN and measure 
their physical properties. Unfortunately it is not always easy to separate extinction 
from chromatic microlensing when variations in color are observed. 

One successful approach to studying chromatic microlensing is the use of 
narrowband photometry (Mosquera et al. 2009). Since the broad line emitting regions 
of quasars are generally too large to be microlensed (e.g., Abajas et al. 2002; Bentz et al. 2009), 
offsets should be observed between the flux ratios of adjacent continuum and emission 
line wavelengths when microlensing is present. 
These offsets cannot be accurately measured using broadband photometry, since 
each filter typically contains contributions from both emission regions. Narrowband filters 
can separate the two, and thus distinguish microlensing from extinction.

In this work, we examine the wavelength-dependent flux ratios of 
SDSS~J1650+4251, HE~0435$-$1223, FBQ 0951+2635, 
and Q~0142$-$100 measured at eight different wavelengths and at several epochs 
using the Nordic Optical Telescope (NOT). These systems were selected because they have previously 
shown evidence for extinction or achromatic microlensing. The details of the observations 
and data reduction techniques are described in Section 2. The modeling and analysis of each individual 
system appears in Section 3. The conclusions and the main results are 
summarized in Section 4.

\section{Observations and analysis}

We observed the gravitational lens systems SDSS~J1650+4251, HE~0435$-$1223, 
FBQ~0951+2635, and Q~0142$-$100 with the 2.56 m NOT 
located at Roque de Los Muchachos, La Palma, Spain.  The images were taken 
with the 2048 $\times$ 2048 ALSFOC detector, which has a spatial scale of 
0.188 arcsec/pixel. We used seven narrowband filters, plus Bessel-I, 
which cover the wavelength interval 3510-8130 \AA. In the case of SDSS~J1650+4251, 
a second round of monitoring observations
was performed using the Liverpool Telescope (LT) with the 2048 $\times$ 2048 RATCam CCD, 
the scale of which is 0.135 arcsec/pixel. We used the Sloan Digital Sky Survey (SDSS) 
ugriz filters covering the wavelength range from 3000 \AA \  to 10000 \AA. Tables \ref{log_NOT} and \ref{log_LT} 
provide a log of our observations and describe the filters. The data were reduced 
using standard IRAF procedures. After overscan and bias correction, a master flat-field 
image was created for each filter and then applied. Finally, frames of the same object 
were aligned and combined for each filter. Cosmic rays were removed when combining the 
frames by using a 3$\sigma$ {\it crreject} rejection algorithm. 

We used {\it imfitfits} (McLeod et al. 1998; Leh\'{ar} et al. 2000) to 
model the images and derive the magnitude differences between the 
quasar images as a function of wavelength. The quasar images were fit as point 
sources and the galaxy luminosity profile was described either with a de Vaucouleurs model or 
an exponential disk. The model consisted of the positions and intensities of all the 
components, as well as the parameters describing the lens galaxy morphology 
(effective/scale radius, ellipticity, and position angle). As these parameters 
were allowed to vary, the models were convolved with several point-spread functions (PSFs; typically three or four) 
defined by stars near each lens system and optimized by computing the $\chi^2$ to fit the observed image. 
If Hubble Space Telescope (HST) data were available, the component positions and the structure 
of the lens galaxy were fixed to the values derived from the HST data.

Since the main contribution to the residuals is caused by systematic errors arising from 
the PSF models rather than statistical fluctuations, we estimated the final magnitude differences 
from the model whose PSF best fits the combined frame. The final uncertainties in the magnitude 
differences were calculated by adding in quadrature two contributions: one that comes 
from the dispersion of the results from the PSFs and the other from the 
dispersion of the fits to the individual frames. In all cases, the residuals 
were negligible compared to the image fluxes. Tables \ref{LT_phot_1650} 
and \ref{NOT_phot_1650} report the results, and the details of the individual models are 
discussed in the next section.

\section{The Individual Systems}

\subsection{SDSS~J1650+4251}

SDSS~J1650+4251 is a two image lens discovered by Morgan et al. (2003). It has a source redshift of $z_{s}= 1.547$. The 
separation of the two QSO components is $1\farcs2$ with B-band magnitudes of 17.8 and 20.0. The
detection of MgII and FeII absorption lines in the spectra of the lensed quasar 
suggests a lens redshift of $z_{L}=0.58$ (Morgan et al. 2003), although
a resolved spectrum of the lensing galaxy is needed to confirm this value.
The time delay for the system is estimated to be $\sim 1$ month (Morgan et al. 2003; 
Vuissoz et al. 2007).

The first series of observations of SDSS~J1650+4251 were obtained with 
the NOT telescope using eight mostly narrowband filters. Due to the good seeing conditions, 
the two quasar components A and B (the brightest and the faintest, respectively) were clearly 
resolved. An unassociated galaxy 4 arcsec south of image A was also detected. To estimate the 
I-band magnitudes, we fit the two components as point sources and modeled the lens galaxy and the nearby ``external'' galaxy as either 
a de Vaucouleurs model or an exponential disk. Since the lens galaxy is very faint, 
it is difficult to distinguish between these two luminosity profiles. 
The best fit was obtained with a de Vaucouleurs profile for the lens galaxy 
and an exponential disk for the nearby galaxy. The excellent seeing conditions 
($0\farcs6$) allow us to determine the relative positions and 
brightnesses of the quasar images, as well as the position of the lens galaxy 
and its flux. The results for the relative component positions in 
Table \ref {pos_1650} are in agreement with the ones obtained by Morgan et al. (2003). 
The broad I-band filter is the only one in which the lens 
galaxy is well detected, appearing as a clear residual if we use a model excluding it.  

\begin{figure}
\begin{tabular}{ccc}
\tabletypesize{\small}

\includegraphics[width=4.5 cm]{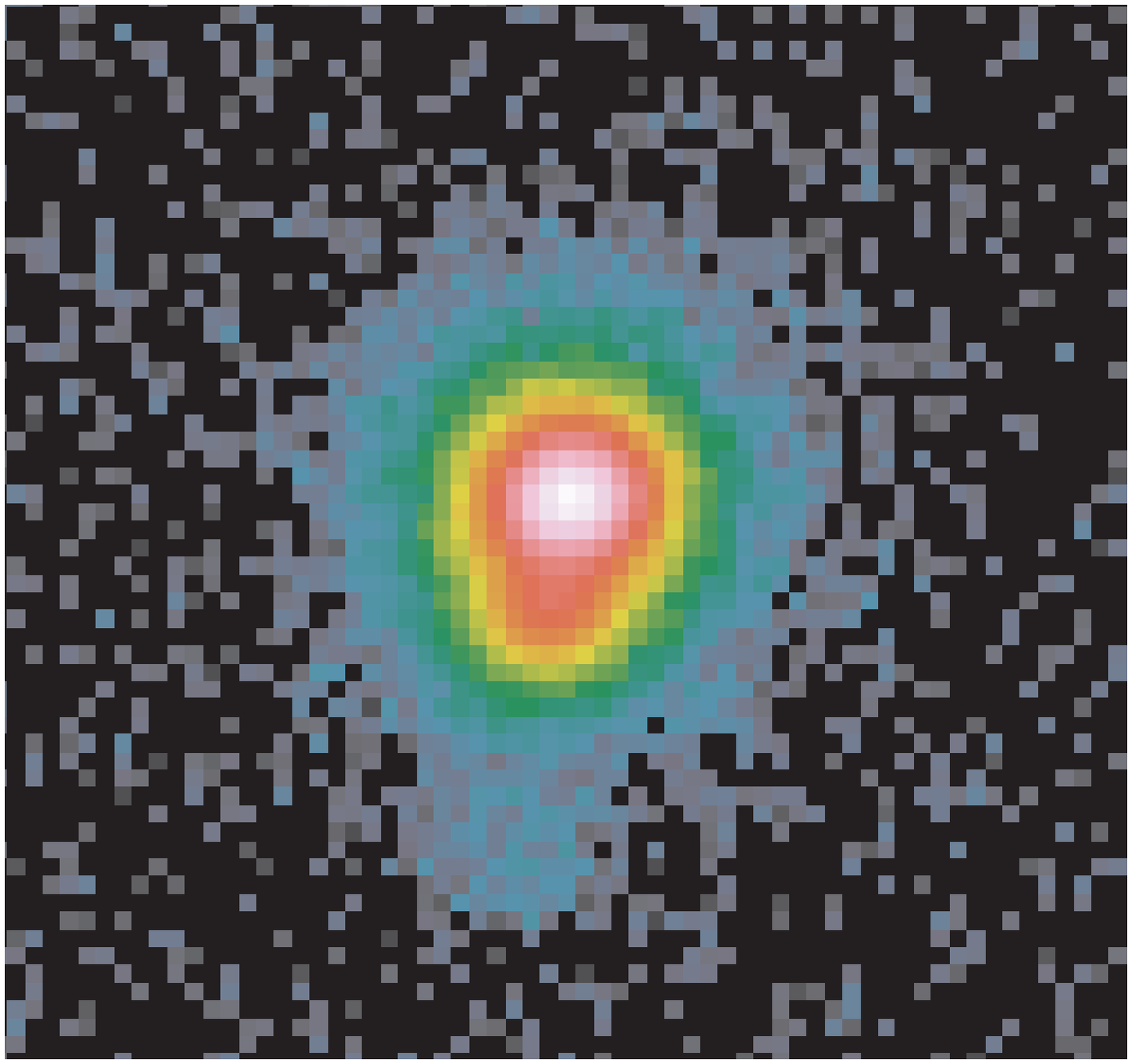} &\includegraphics[width=4.5 cm]{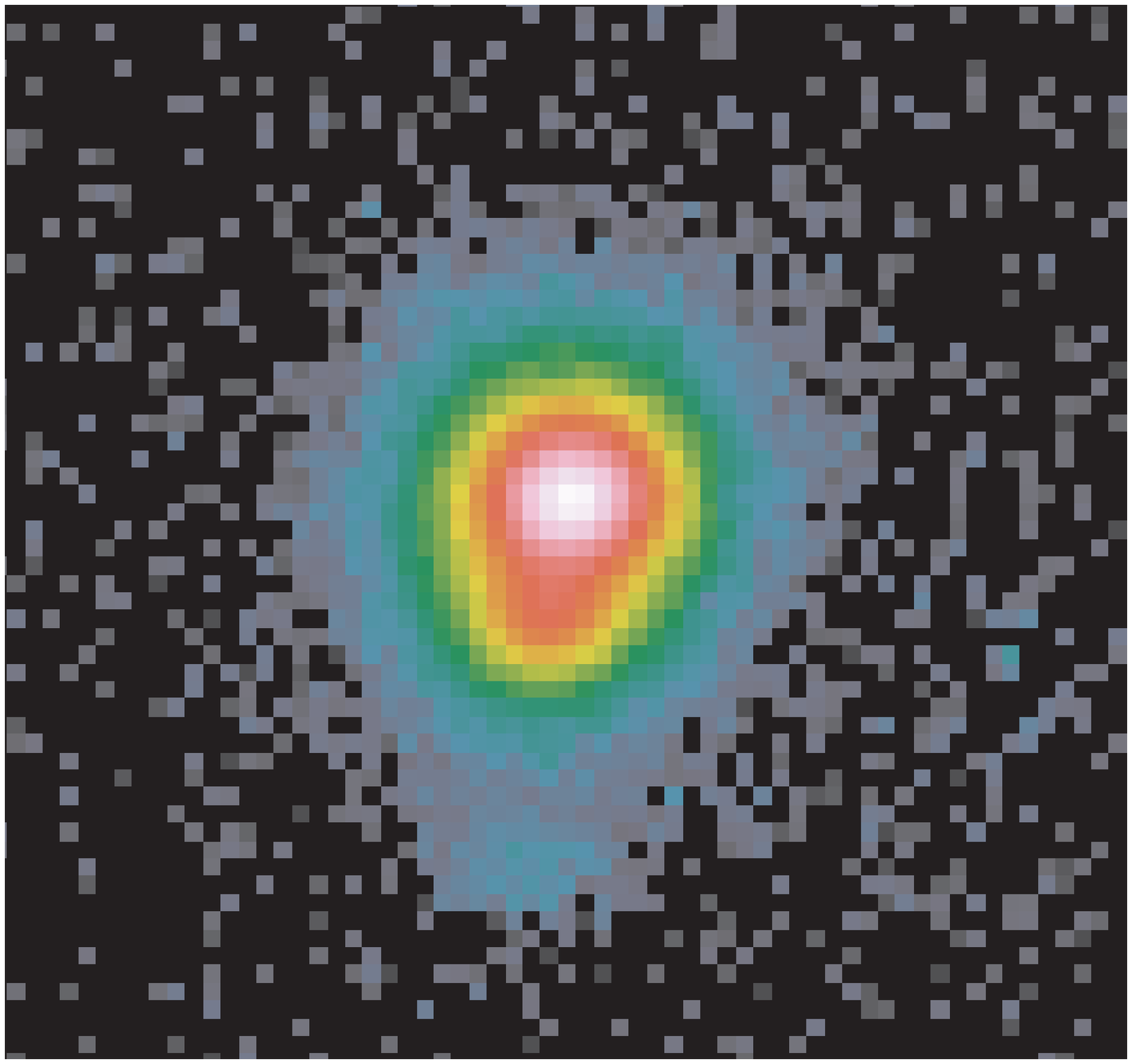}&
\includegraphics[width=4.5 cm]{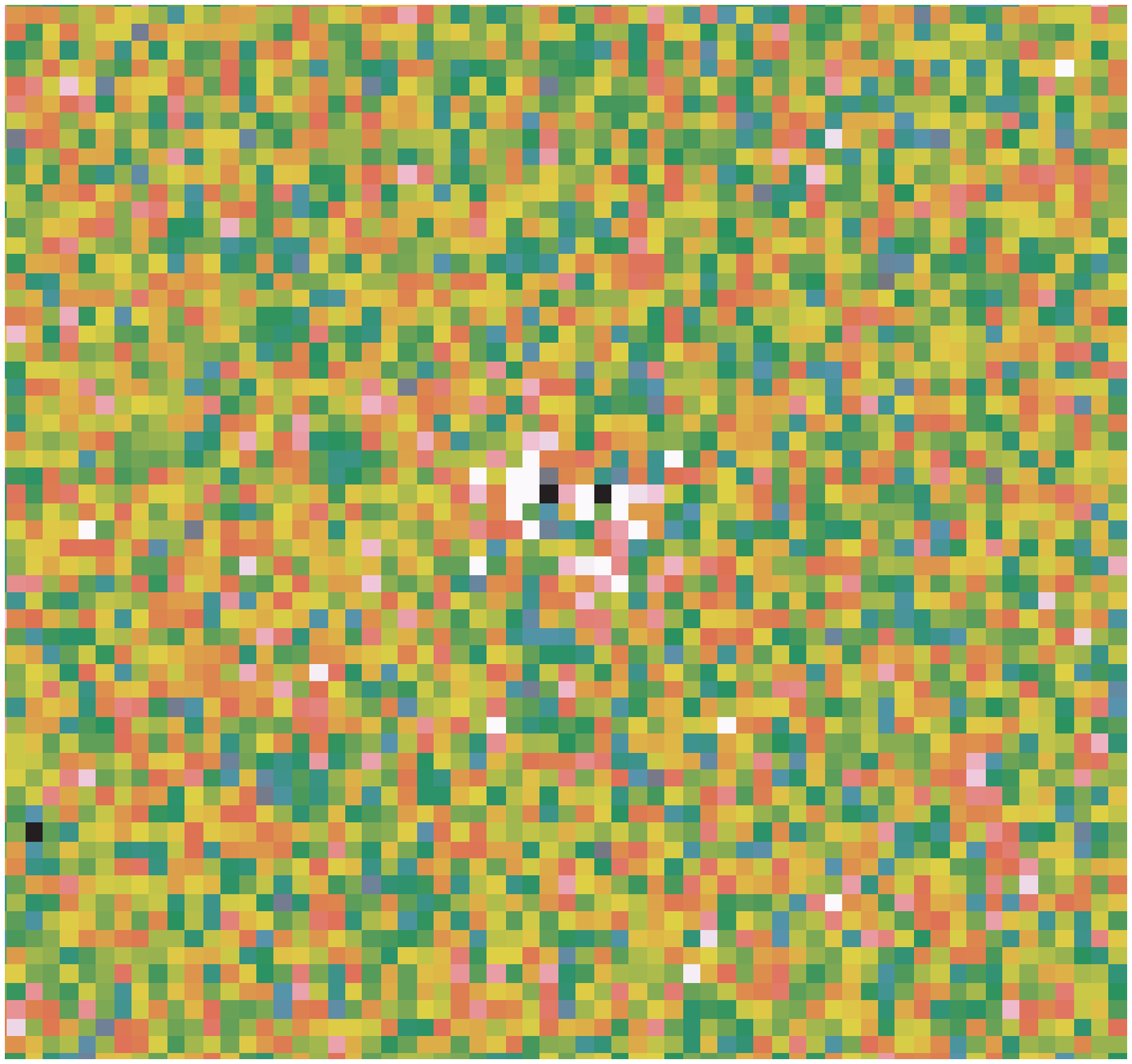}\\
SDSS~J1650+4251 & Modelled image & Residuals\\
\end{tabular}
\caption{\label{modelo_y} \small Str-y image (left) of SDSS~J1650+4251, our photometric model 
consisting of two point sources (middle), and the residuals (right) after subtracting the model from 
the data. The peak residuals (white) are roughly 2\% of the quasar peak.}

\end{figure}

\begin{figure}[t]
\begin{center}
\vspace{0.5 cm}
\includegraphics[scale=0.5]{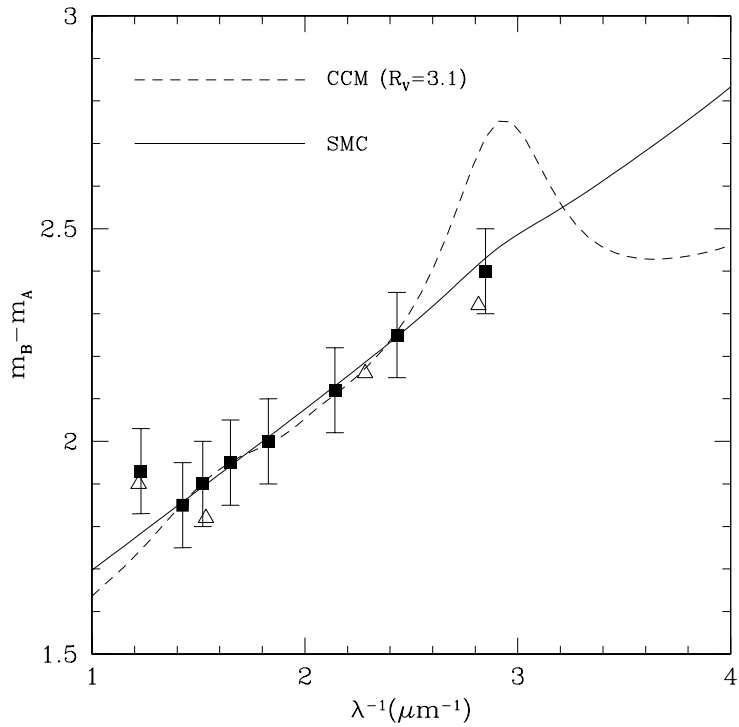}
\caption{\label{1650_extinction} \small Difference in magnitude vs. the 
inverse of the observed wavelength for the
quasar images of the lens system SDSS~J1650+4251. The filled squares correspond to our 
NOT data and the open triangles to those obtained by Morgan et al. (2003). The 
variation in the magnitude difference with wavelength is probably produced by differential 
extinction in the lens galaxy. An extinction law similar to the SMC fits the observations 
well, while a Milky Way extinction law (CCM) is ruled out.}
\end{center}
\end{figure}

In the seven narrowband filters, the contribution from the lens galaxy is very small, 
and the lens system can be modeled simply as two point sources.
The resulting fits are excellent, as illustrated by the residuals for the Str\"omgren-y image 
shown in Figure \ref{modelo_y}. The peak residuals for the images are less than  
1.7\%  of the peak intensity of image A. The absence of the lens galaxy in the residuals, 
even at wavelengths beyond the 4000 \AA~break, is consistent with the expected level of contamination. 
Assuming that the calculated flux ratios (with and without galaxy) are well determined 
in the I-band, we can estimate the contamination in the bluer filters using a quasar spectrum template 
for the images and an early-type spectrum template for the lens galaxy (Assef et al. 2010).  We find that 
the lens galaxy contamination in the flux ratios 
would be at most $\sim 0.04$ mag in the bluer filters, within the level of the model uncertainties. 
The photometric results appear 
in Table \ref{NOT_phot_1650} and are plotted as a function of wavelength in 
Fig. \ref{1650_extinction} (black squares). In Figure \ref{1650_extinction} we have also 
included the results of Morgan et al. (2003) (empty triangles), which are in good agreement 
with the ones we obtained.

To model the effects of differential extinction we fit the magnitude differences as 
a function of wavelength following Falco et al. (1999) as
\begin{equation}\label{ext_AB}
m_{B}(\lambda)- m_{A}(\lambda)= \Delta M + \Delta E \ R 
\left(\frac{\lambda}{1+z_{l}}\right),
\end{equation}
where $\Delta M=M_{B}-M_{A}$ is the relative magnification, $\Delta 
E(B-V)=E_{B}(B-V)-E_{A}(B-V)$ is the differential extinction, 
and $R(\lambda)$ is the mean extinction law. The fits can also be done as 
a function of the  ``dust redshift'' (Jean $\&$ Surdej 1998), either to estimate the lens 
redshift or as an added confirmation of extinction. All these quantities 
can be determined without needing to know the intrinsic spectrum of the quasar, $m_{0}(\lambda)$.
We use a $\chi^2$ statistic for the fits and either the CCM (1989) parameterized 
models for the Galactic extinction curve or the Fitzpatrick $\&$ Massa (1990) 
model with its parameters set to the values found by Gordon et al. (2003) 
for the average extinction in the SMC. The results are shown in 
Figure \ref{1650_extinction}.

Our best fit for the data with $z_{dust}=0.58$ is obtained for an SMC extinction law
with  $\chi^2_{dof}=0.08$\footnote{We neglect the somewhat offset I-band 
point. We think that it is discrepant because the lens galaxy was 
slightly oversubtracted.}, where $\chi^2_{dof}$ is  $\chi^2$ per degree of freedom. 
The parameters for this fit are $\Delta M= 1.8\pm 0.1$ and $\Delta E= 0.10\pm0.02$ at 
1$\sigma$. For the CCM extinction model with $R_{V}=3.1$, we find $\chi^2_{dof}=2.2$, which is a 
significantly worse fit because the Str-u point is in conflict with the presence 
of a 2175 \AA~ feature. Excluding the Str-u point, the CCM model would also fit well ($\chi^2_{dof}=0.03$).

\begin{figure}[t]
\begin{center}
\vspace{0.5 cm}
\includegraphics[scale=0.5]{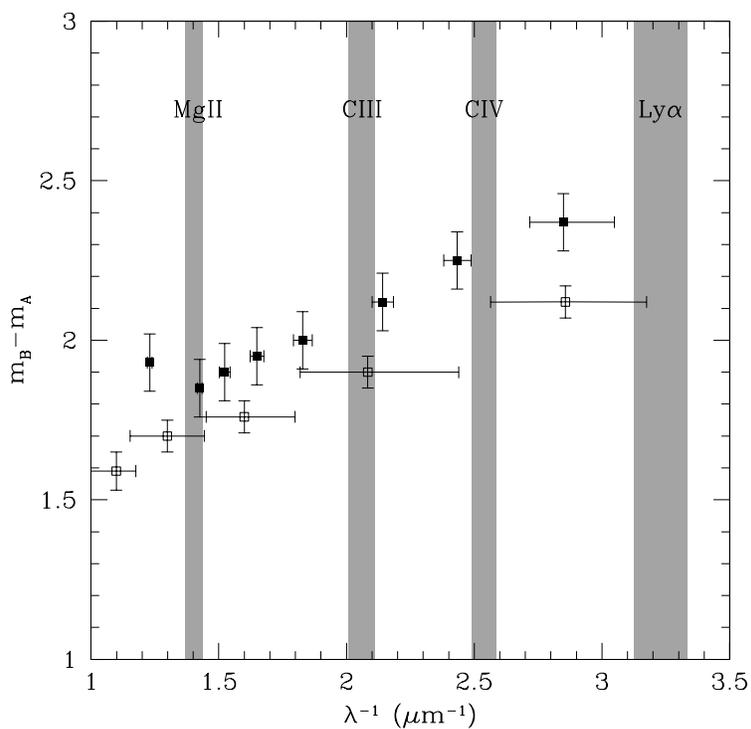}
\caption{ \label{LT_1650} \small Difference in magnitude vs. the 
inverse of the observed wavelength for the
quasar images of the lens system SDSS~1650+4251. The filled squares 
correspond to the data taken at the NOT (2003) and the open squares correspond 
to data taken at the LT (2006). The trend with wavelength  is the same in all the different epochs, 
although a magnitude shift due to microlensing may also be present. The wavelength regions 
corresponding to the most prominent quasar emission lines are indicated (from 
left to right: MgII, CIII, and CIV ). The horizontal error bars correspond 
to the filter widths.}
\end{center}
\end{figure}

The alternative to extinction as an explanation for the wavelength-dependent flux 
ratios is chromatic microlensing. We view this as a less likely explanation 
because the wavelength dependence observed in the NOT data is consistent  with that 
observed  $\sim 4$ months earlier by Morgan et al. (2003) 
and  $\sim 3$ years later in our LT observations, 
as shown in Figure \ref{LT_1650}. There is a wavelength-independent 
shift of $\sim 0.2$ mag that could  be due to microlensing, 
but the chromatic structure is unchanged. Since the 
observations made at the LT cover a period larger than the expected time delay and no 
significant variations were observed between the different nights, the magnitude 
shifts are probably not due to intrinsic variability in the quasar modulated by the time delay.

\subsection{HE~0435$-$1223}

HE~0435$-$1223 was discovered by Wisotzki et al. (2000) in the 
Hamburg/ESO survey for bright QSOs and later identified as a gravitational lens 
(Wisotzki et al. 2002). In this quadruple system, a 
background quasar with a redshift of $z_S=1.689$ is gravitationally lensed 
by an early-type galaxy at $z_L=0.4541$ (Morgan et al. 2005). Integral-field 
spectrophotometry  (Wisotzki et al. 2003) showed no evidence for differential 
extinction, while there was evidence for microlensing. Kochanek et al. (2006) 
also observed microlensing variations $\sim 0.1$ mag yr$^{-1}$. Due to 
the symmetric distribution of the images around the lensing galaxy, the time delays of  
$\Delta t_{AB}=-8.00^{+0.73}_{-0.82}$ days, 
$\Delta t_{AC}= -2.10^{+0.78}_{-0.71}$ days and $\Delta t_{AD}= -14.37^{+0.75}_{-0.85}$ 
days (Kochanek et al. 2006) are relatively small. Courbin et al. (2010) found 
similar time delays using a longer monitoring period but with larger formal uncertainties.

\begin{figure}
\begin{center}
\includegraphics[scale=0.5]{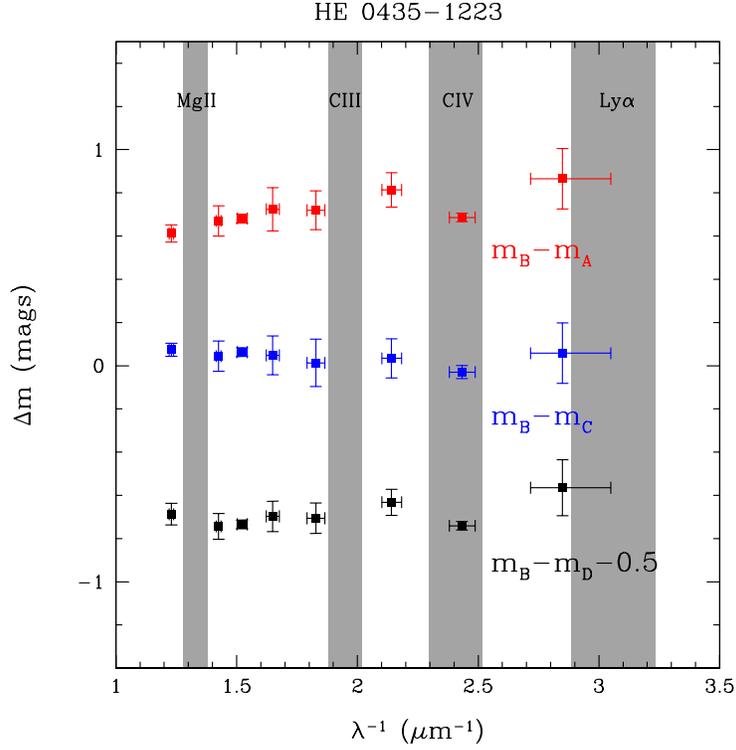}
\caption{ \label{0435_over} \small Magnitude differences as a function of the 
inverse of the observed wavelength 
for HE~0435$-$1223 (NOT data on HJD 2454404). Images A and D show a 
clear signature of microlensing in the continuum but not in the 
emission lines. Moreover, the chromatic variation observed 
in image A could only be explained as a consequence of chromatic microlensing. The wavelength regions 
corresponding to the most prominent quasar emission lines are indicated (from left to right: MgII, CIII, CIV 
and Ly$\alpha$). The horizontal error bars correspond 
to the filter widths. }
\end{center}
\end{figure}

Observations of this system were made using the narrowband filter set at the
NOT on two nights separated 15 days (see Table \ref{log_NOT}), and the 
photometric results for the two epochs agree at the 1-$\sigma$ level. Figure \ref{0435_over} 
shows the results of the PSF photometry for one of those nights, in which 
the lens galaxy was modeled with a de Vaucouleurs profile, and the quasar images 
were included as point sources. Their relative positions and the structure of the 
lens galaxy were fixed to the HST values 
(Kochanek et al. 2006).

The behavior of the flux differences in Figure \ref{0435_over} 
looks very similar to what we found when studying Q~2237+0305 (Mosquera et al. 2009), 
suggesting that we have detected chromatic microlensing in 
another quadruple system. Considering the quasar redshift, the Str\"omgren-u 
and Str\"omgren-v filters are the only ones affected by the broad 
emission lines of the quasar. The Ly$\alpha$ emission line contributes 
about 38\% of the flux in the Str\"omgren-u filter, and part of the CIV emission 
line extends over the full width of the Str\"omgren-v filter based on 
the SDSS composite quasar spectrum (Vanden Berk et al. 2001). Therefore, in 
a system affected by microlensing, these two filters will be offset from the continuum 
flux ratios at those wavelengths, because the broad line regions are much larger than 
the length scales for microlensing (e.g., Abajas et al. 2002; Bentz et al. 2009).

\begin{figure}[t]
\begin{center}
\includegraphics[scale=0.5]{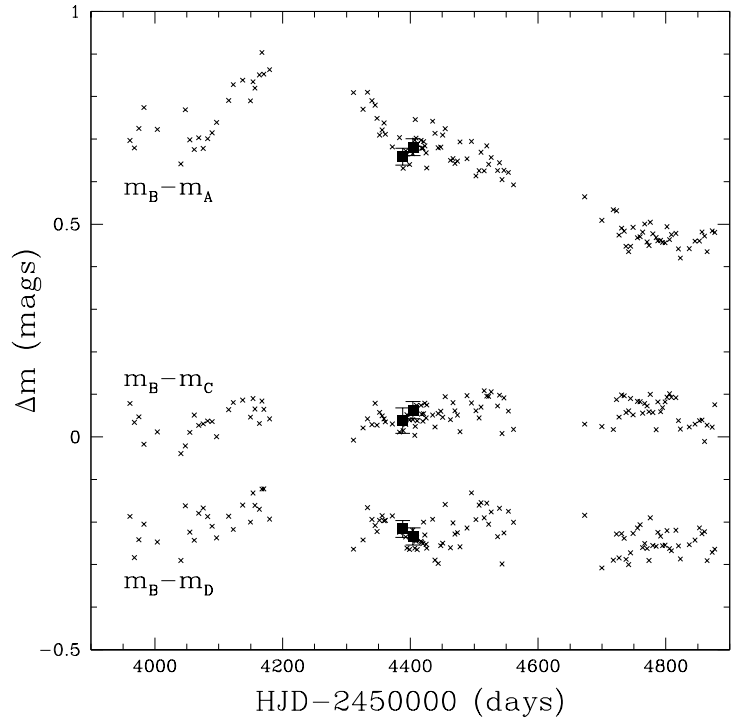}
\caption{ \label{fig:CSK} \small HE~0435$-$1223 SMARTS R-band light 
curve (Blackburne $\&$ Kochanek 2010) corrected for time delays. The filled squares 
correspond to our observation in the redshift zero H$\alpha$ filter, which is a good 
match to the R-band wavelength.}
\end{center}
\end{figure}

Looking  in Figure \ref{0435_over} at the filters that are not affected by emission lines, 
it appears that chromatic microlensing is affecting image A. The observed chromaticity 
between the bluest (Str-b) and the reddest (I-band) filters is $(\Delta m)_{I-b}=0.20\pm0.09$. 
As none of B, C, or D shows significant signs of a wavelength dependence, even at wavelengths 
contaminated by emission lines, achromatic microlensing must be weak for these images. 
This is supported by the R-band light curves obtained with the SMARTS 1.3 telescope located at the Cerro Tololo 
Inter-American Observatory, in Chile (Blackburne $\&$ Kochanek 2010). 
Figure \ref{fig:CSK} shows the brightness fluctuations observed in the 
difference ($m_B-m_A$), once a time delay correction 
of $-8.0$ days was applied, and time-dependent changes are not seen in the 
other two differences ($m_B-m_C \sim 0.05$ and $m_B-m_D \sim -0.25$ ). Thus, image A seems 
to have been undergoing a  microlensing event near the time of 
our NOT observations, while no-microlensing variations were observed in B, C, and D.  
Our flux ratios for the redshift zero H$\alpha$ filter agree well with the similar-wavelength 
SMARTS R-band observation (Figure \ref{fig:CSK}, filled squares).

We modeled the microlensing of image A as follows. First, we fit a simple 
singular isothermal ellipsoid plus external shear model to the HST positions 
from Kochanek et al. (2006) using the lensmodel package (Keeton 2001) to determine a convergence 
and shear for image A of  $\kappa=0.43$ and $\gamma=0.39$. Second, we 
generated microlensing magnification patterns using the inverse polygon 
mapping method of Mediavilla et al. (2006). We generated patterns with stellar 
mass fractions of $\kappa_{\ast}/\kappa= 0.01$, $0.05$, $0.1$, $0.15$, $0.2$, $0.25$. 
We used stellar masses of $M$=1 $M_{\odot}$, an outer scale of $20 \ r_{E}$ and a 
dimension of $2048 \times 2048$ pixels to get a resolution of $0.01$ $r_{E}/$pixel. 
With the present data we cannot determine $M$, and our size estimates can be rescaled as $(M/M_{\odot})^{1/2}$.
Finally, we estimated the R-band microlensing magnification of image A assuming that 
the H-band fluxes from Kochanek et al. (2006) represented the true flux ratios. The H-band data 
are the reddest ones reported in the literature for HE~0435$-$1223, and will be less affected by 
extinction and chromatic microlensing. Even though chromatic effects could be contaminating 
the H-band measurements, our observations support the hypothesis that they are negligible. 
The H-band magnitude differences, $(m_B-m_C)_H= -0.02\pm0.04$ and $(m_B-m_D)_H=-0.23\pm 0.06$, match the 
respective SMARTS R-band differences (see Figure \ref{fig:CSK}). Since from the NOT chromatic data 
we know that neither $B$, $C$, nor $D$ is significantly microlensed, the H-band flux ratios 
should be a good estimate of the real fluxes of the images. This means that at R-band 
image A is magnified by microlensing by $\Delta M_{R}=-0.19\pm0.04$. The differential microlensing of image A 
between the Str-b and R bands was simply estimated  
from the observed chromatic microlensing, and is 
$\Delta M_{b-R}=-0.13\pm0.08$. 

\begin{figure}
\begin{center}
\includegraphics[scale=0.8]{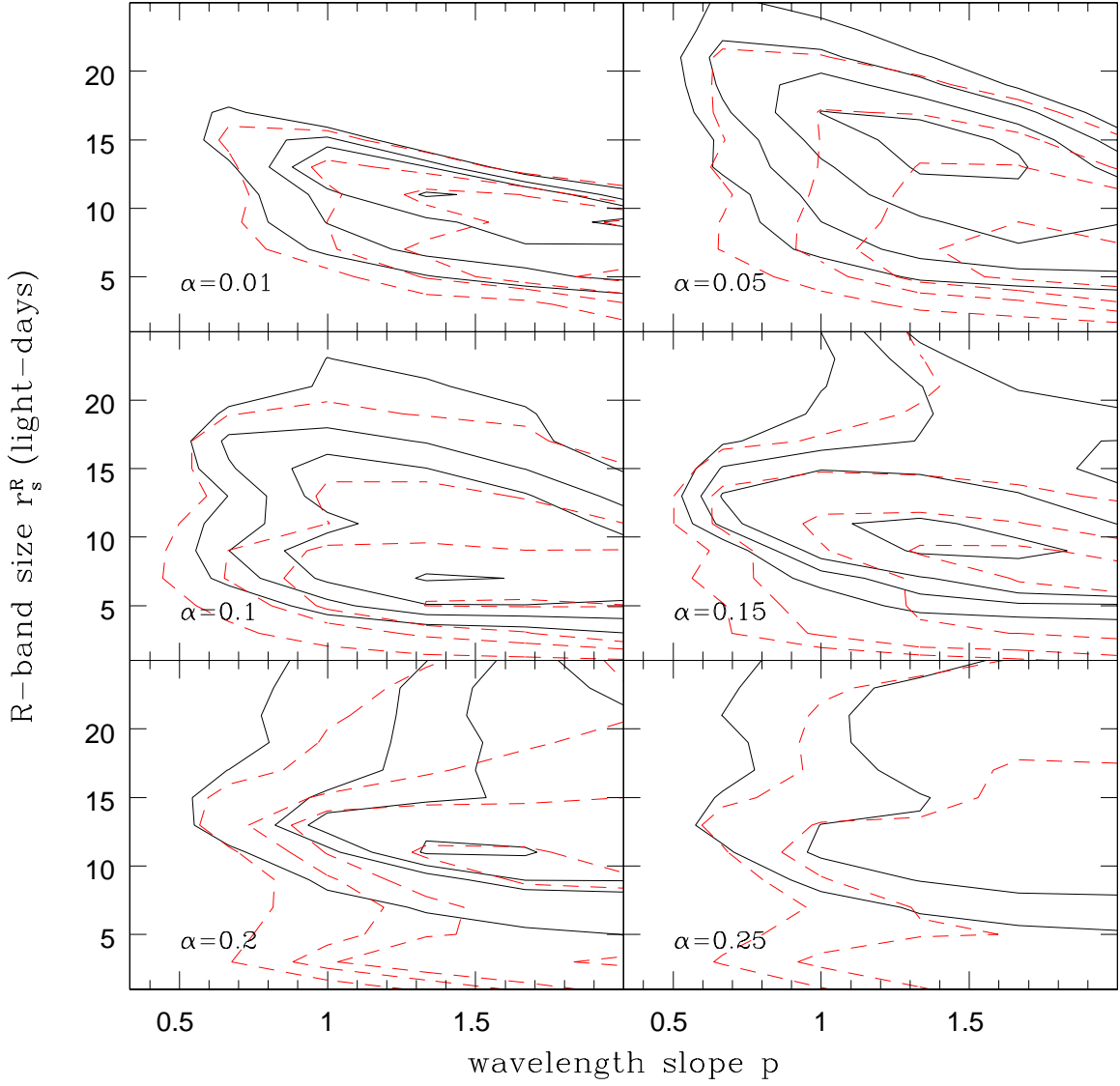}
\caption{ \label{probs_rp} \small Probability of reproducing the observed microlensing 
magnification and chromaticity of HE~0435$-$1223 as a function of the R-band disk size, 
$r^R_s$, and the power-law index, $p$, where $r^R_s\propto \lambda^p$. The panels show the results for different stellar 
mass fractions $\alpha$. The contours correspond to 15\%, 47\%, 68\%, and 90\% confidence 
intervals. Uniform (solid line) and logarithmic (dashed line) priors on $r^R_s$ were assumed.}
\end{center}
\end{figure}

\begin{figure}
\begin{center}
\includegraphics[scale=0.8]{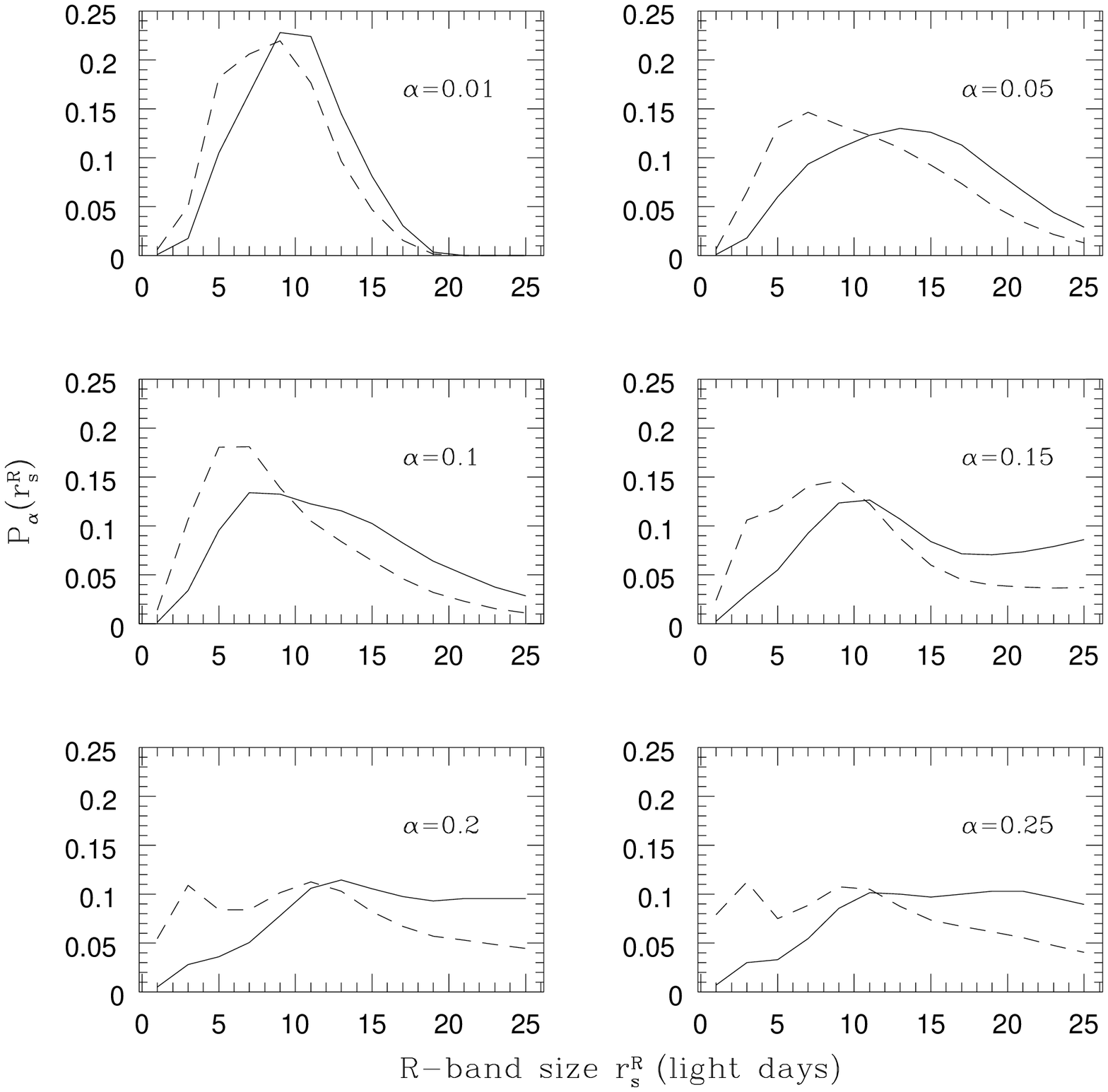}
\caption{\label{probs_r} \small Probability of reproducing the observed microlensing 
magnification and chromaticity of HE~0435$-$1223 as a function of the R-band disk size, $r^R_s$ 
for different stellar mass fractions $\alpha$, and normalized by the total probability in 
the ($r^R_s$, $p$) grid. Uniform (solid line) and 
logarithmic (dashed line) priors on $r^R_s$ were assumed.}
\end{center}
\end{figure}

\begin{figure}
\begin{center}
\includegraphics[scale=0.8]{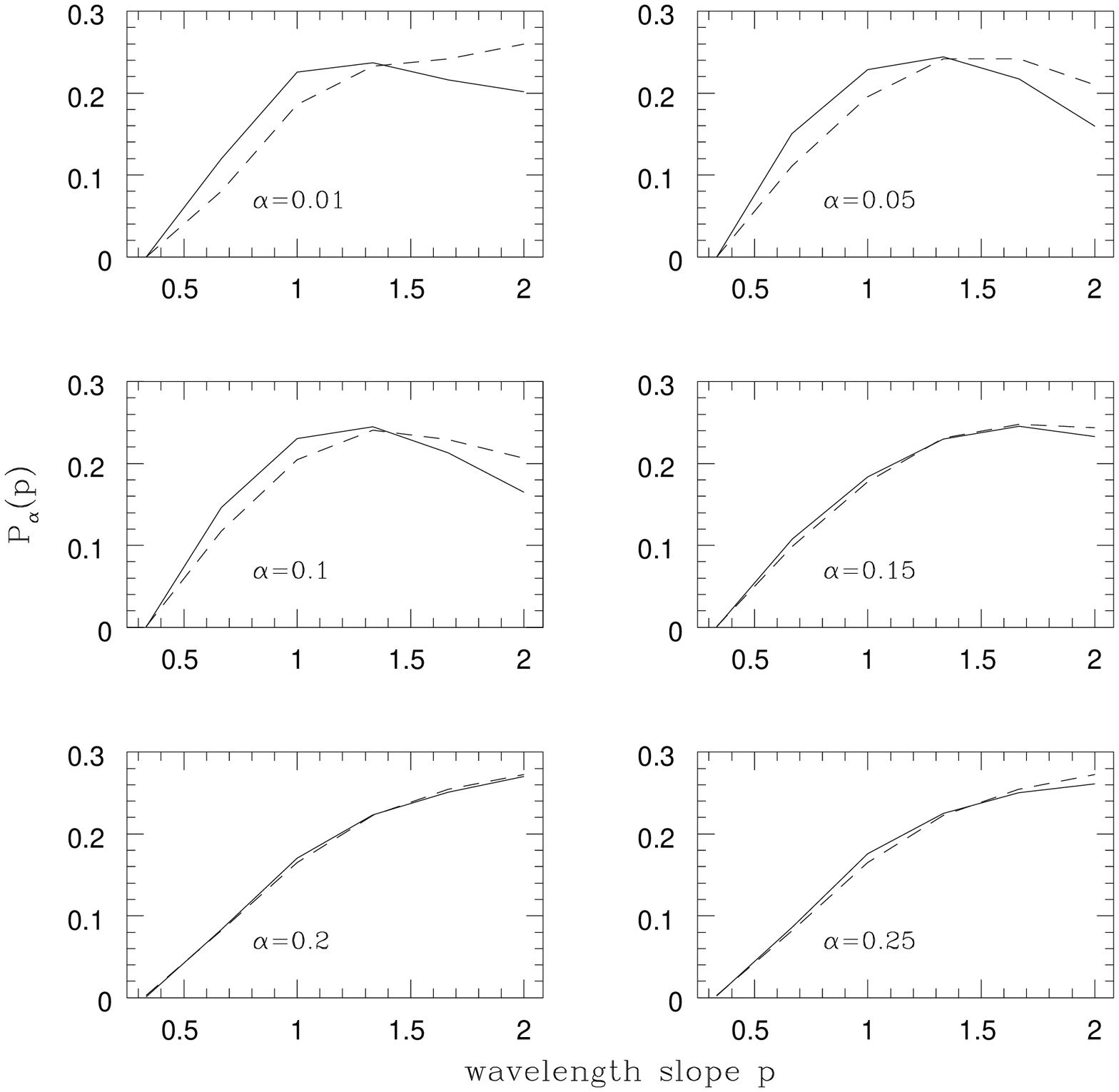}
\caption{\label{probs_p} \small Probability of reproducing the observed microlensing 
magnification and chromaticity of HE~0435$-$1223 as a function of the power-law index, $p$ 
for different stellar mass fractions $\alpha$, and normalized by the total probability in 
the ($r^R_s$, $p$) grid. Uniform (solid line) and logarithmic 
(dashed line) priors on $r^R_s$ were assumed.}
\end{center}
\end{figure}

\begin{figure}[t]
\begin{center}
\includegraphics[scale=0.5]{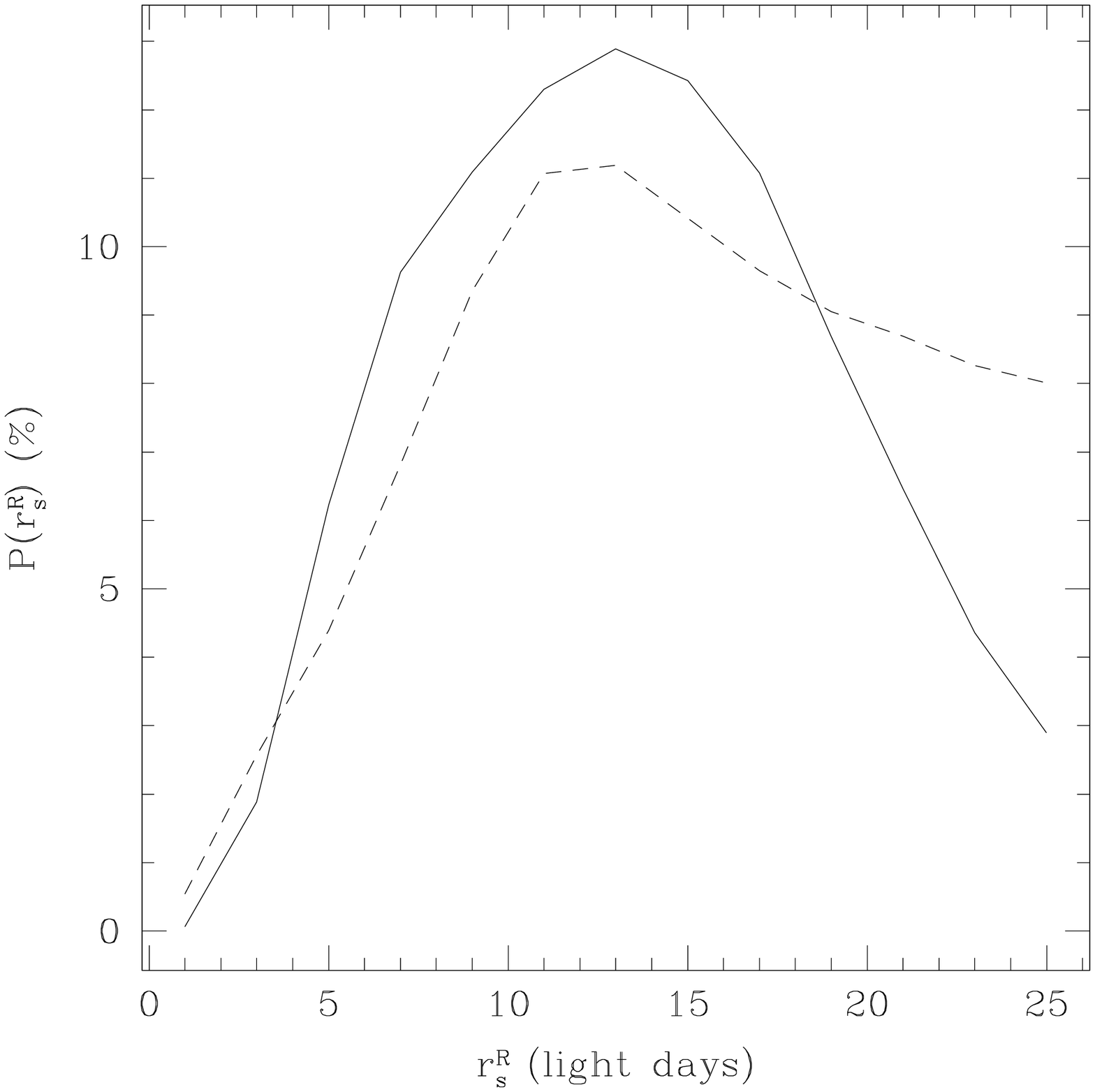}
\caption{ \label{Rint} \small Probability of reproducing the observed microlensing 
magnification and chromaticity in HE~0435$-$1223 as a function of the R-band disk size, $r^R_s$, after
marginalizing over $\alpha$. The solid line was obtained using the prior on $\alpha$ from Mediavilla et al. 
(2009). The dashed line corresponds to the probability distribution without this prior. These results 
are for the uniform prior on $r^R_s$.}
\end{center}
\end{figure}

\begin{figure}[t]
\begin{center}
\includegraphics[scale=0.5]{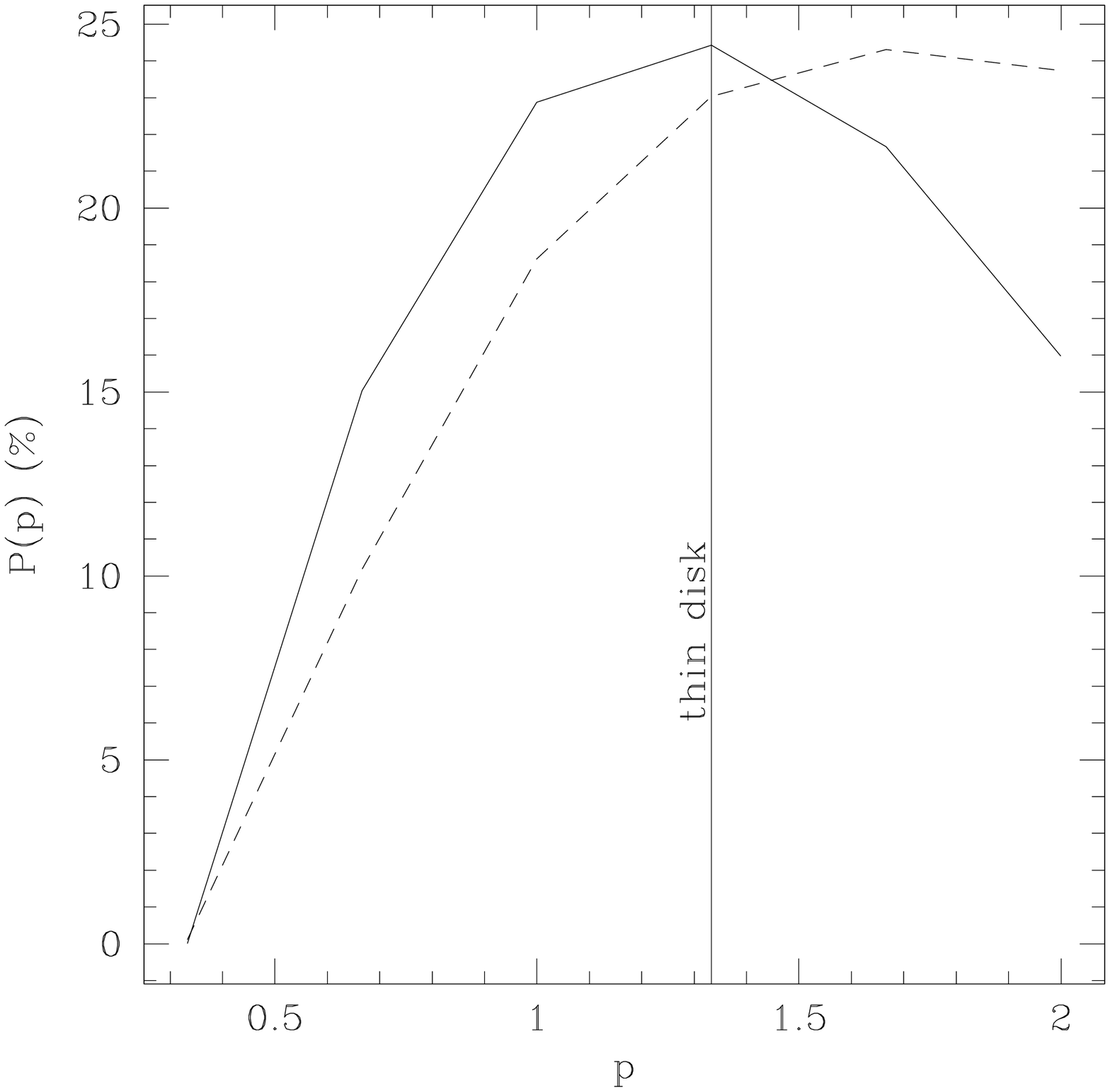}
\caption{ \label{pint} \small Probability of reproducing the observed microlensing 
magnification and chromaticity in HE~0435$-$1223 as a function of the power-law index, $p$, after 
marginalizing over $\alpha$. The solid line was obtained using the prior on $\alpha$ from Mediavilla et al. 
(2009). The dashed line corresponds to the probability distribution without this prior. These results 
are for the uniform prior on $r^R_s$.}
\end{center}
\end{figure}

Finally, we convolved the magnification patterns with Gaussian source 
profiles ($\propto \exp(-r/2r_s)$) of varying size $r^{R}_s$ to model the R-band  and  
$r^{b}_s= r^{R}_s (\lambda_b/\lambda_R)^{p}$ for the Str-b band to 
compute the probability of reproducing $\Delta M_{R}$ and $\Delta M_{b-R}$ as a function 
of the size $r^{R}_s$, the power-law index $p$, and the stellar fraction  $\kappa_{\ast}/\kappa$ 
(see Mosquera et al. (2009) for details on the probability distribution calculations). We adopted 
$p=\frac{i}{3}$, $i=$1, ..., 6, and $r^{R}_s=(1+2i)$ light-days, 
$i=$0, ..., 12. Figure \ref{probs_rp} (solid lines) shows the probability 
distribution $P(r^{R}_s, p)$ for the different values of the stellar fraction $\alpha=\kappa_{\ast}/\kappa$. 
The contours correspond to 15\%, 47\%, 68\%, and 90\% confidence intervals. The integrated values of the 
probabilities, $P_{\alpha}(r^{R}_s)$ and $P_{\alpha}(p)$, are shown in 
Figures \ref{probs_r} and \ref{probs_p}, respectively, 
for the considered stellar fractions. If we adopt the likelihood maximum as our estimator for $r^{R}_s$ 
and $p$, they can be constrained at 68\% confidence for $\alpha<0.15$ (Table \ref {tab_Rpa}). 
For $\alpha>0.15$ the shape of the probability distributions does not allow us to 
determine the corresponding uncertainties. However, we know from Mediavilla et al. (2009) and from time 
delay measurements (Kochanek et al. 2006) that low values of $\kappa_{\ast}/\kappa\lesssim0.2$ are favored. 
Therefore we calculated $P(r^{R}_s)$ and $P(p)$ using the probability distribution of the stellar mass fraction 
found by Mediavilla et al. (2009) as a prior (solid line in Figures \ref{Rint} and \ref{pint}), 
interpolating over the probability distributions since the sampling in $\alpha$ is not uniform. From these weighted 
probability distributions we estimate that the disk size and the power-law slope are $r^{R}_s=13\pm5$ 
light days and $p=1.3\pm0.3$ at 68\% confidence. We also estimated these values for a logarithmic prior on 
$r^{R}_s$ (Table \ref {tab_Rpa}) just by dividing by $r^{R}_s$ the distributions obtained with a uniform prior. The probability 
distributions $P(r^{R}_s, p)$ using this prior, and the corresponding integrated values $P_{\alpha}(r^{R}_s)$ 
and $P_{\alpha}(p)$, are also shown in Figures \ref{probs_rp}, \ref{probs_r}, 
and \ref{probs_p}, respectively (dashed lines). In this case we found a disk size 
of  $r^{R}_s=7\pm6$ light days and a similar slope $p$. Pooley et al. (2007, 2009) 
and Blackburne et al. (2011) found similar disk sizes to these estimates by comparing optical and X-ray flux 
ratios, and similar results were found by Morgan et al. (2008, 2010) modeling the R-band light curves (they 
estimated $r^{R}_s=15^{+23}_{-9}$ light days for a face-on quasar). The slope $p$ for the 
wavelength dependence is in agreement with the Shakura $\&$ Sunyaev (1973) simple disk model ($p=4/3$). 
Blackburne et al. (2011) also found solutions compatible with our results for HE~0435$-$1223 
with $p=0.55\pm0.49$ and $p=0.67\pm0.55$ for a linear and for a logarithmic prior, respectively, although 
they found a shallower average slope for their full sample of lenses. 

\begin{figure}[t]
\begin{center}
\includegraphics[scale=0.5]{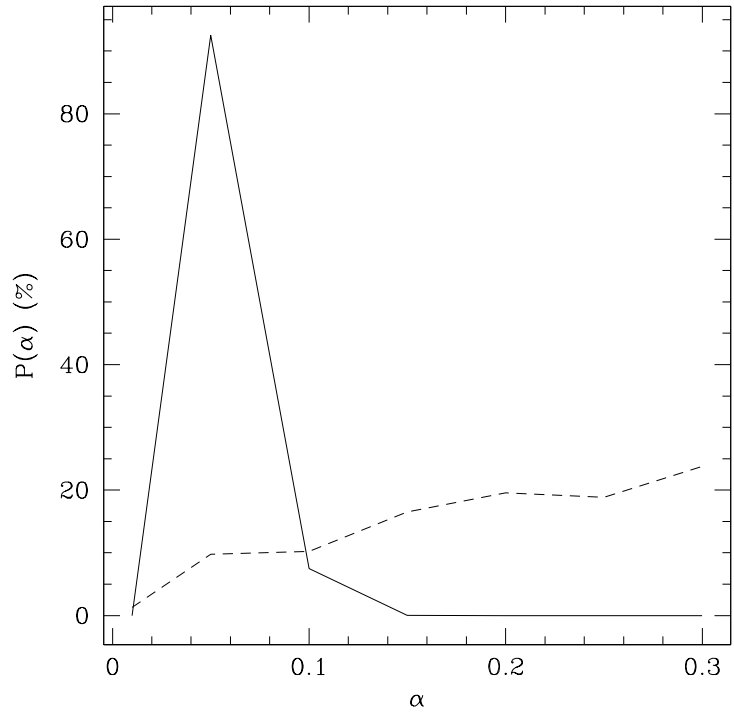}
\caption{ \label{alphai} \small Probability of reproducing the observed microlensing 
magnification and chromaticity HE~0435$-$1223 as a function of the stellar fraction, 
$\alpha$, marginalized over the R-band disk size $r^R_s$ and the power-law index $p$. 
The solid line was obtained using the prior on $\alpha$ from Mediavilla et al. 
(2009). The dashed line corresponds to the probability distribution without this prior. These 
results are for the uniform prior on $r^R_s$.}
\end{center}
\end{figure}

If we compare the probability distribution $P(\alpha)$ without the Mediavilla et al. (2009) prior, 
it increases monotonically with $\alpha$ (Figure \ref{alphai}, dashed line). This is the expected 
behavior for $P(\alpha)$, since a single epoch chromatic microlensing detection likely introduces 
a strong bias toward large $\alpha$  values. With $\alpha$ large there are many more regions with 
the strong micro-magnification gradients needed to produce the observed chromaticity. This is the 
reason why we cannot constrain the stellar fraction with our procedure. When we include the 
Mediavilla et al. (2009) prior, the resulting $P(\alpha)$ resembles that prior (Figure \ref{alphai}, 
solid line). In the end, our procedure is essentially equivalent to computing the probability distributions 
$P(r^{R}_s)$ and $P(p)$ for a fixed value of $\alpha\simeq0.1$ that is also consistent with the time delays.

\subsection {FBQ~0951+2635}

\begin{figure}
\begin{center}
\vspace{0.5 cm}
\includegraphics[scale=0.5]{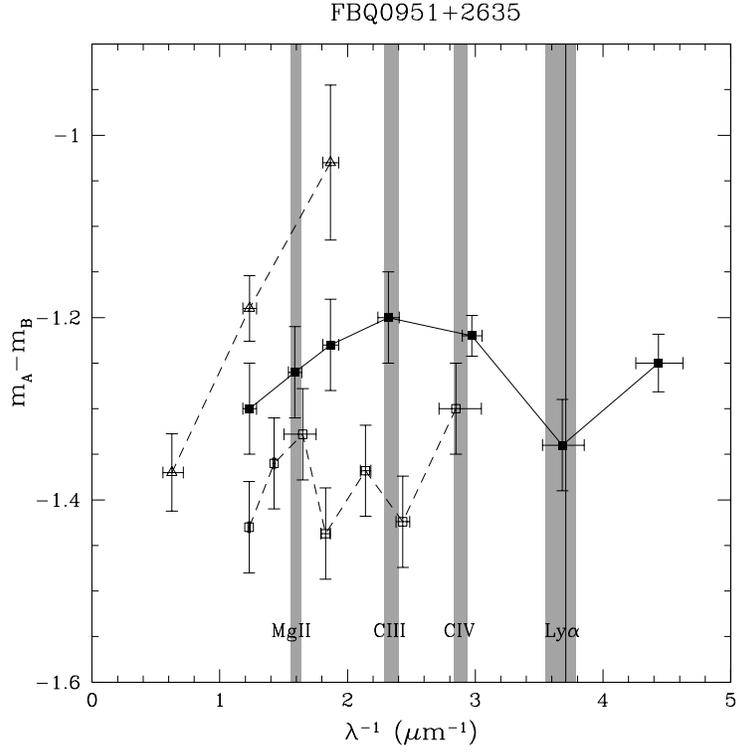}
\caption{\label{0951_phot} \small
Magnitude differences as a function of the inverse of the observed wavelength 
for FBQ~0951+2635. The NOT observations (open squares) are compared with the 
CASTLES (open triangles) and J. A. Mu\~noz et al. (2011, in preparation, filled squares) HST observations. 
The wavelengths corresponding to the most prominent quasar emission lines are indicated 
(from left to right: MgII, CIII, CIV and Ly$\alpha$). The horizontal error bars correspond 
to the filter widths. The solid vertical line in the Ly$\alpha$ band corresponds to the 
expected position of the 2175 \AA~feature.}
\end{center}
\end{figure}

The gravitational lens system FBQ~0951+2635 was discovered by Schechter et al. (1998). It has two quasar 
images separated by $1\farcs1$, 
and the time delay between them is estimated to be approximately two weeks (e.g., Jakobsson et al. 2005). 
The quasar redshift is $z_S=1.246$. However the redshift of 
the lens galaxy was much more difficult to measure, since it was very difficult to disentangle
its flux from the quasar images. Kochanek et al. (2000) suggested a value
of $z_L=0.21$ from the position of the lens in the fundamental plane, and this 
was spectroscopically confirmed to be $z_L=0.260 \pm 0.002$ by Eigenbrod et al. (2007).
Indications of microlensing in the system were found by several
authors (e.g., Schechter et al. 1998; Jakobsson et al. 2005), but the 
chromatic behavior of FBQ~0951+2635 is still not well understood. This system was 
observed during one night at the NOT (see Table \ref{log_NOT}) using 
our narrowband filter set. We fit the data using point sources for the quasar 
images, and the lens galaxy was included in the I-band model as a de 
Vaucouleurs profile. Figure \ref{0951_phot} shows the result of our photometry compared 
with HST measurements at two different epochs (CASTLES and J. A. Mu\~noz et al. 2011, in preparation).

If we focus on the NOT data in Figure \ref{0951_phot} (open squares), our results are 
compatible with no chromaticity. However, our data shed little light on the chromatic 
behavior of FBQ~0951+2635, because almost all the filters contain strong emission lines. 
The iac\#29 filter corresponds to the OIII line, iac\#28 is blended with MgII, 
the [NeIV], FeIII and [OII] lines lie in the Str-y filter, Str-v is affected at 
$\sim50$ \% by CIII], and the Str-u filter contains SiIV, OIV], CIV, and HeII. In any 
case, our results as well as those obtained by J. A. Mu\~noz et al. (2011, in preparation; 
Figure \ref{0951_phot}), confirm the lack of extinction because no chromatic 
fluctuations are observed. The J. A. Mu\~noz et al. (2011, in preparation) flux 
ratios are shifted by $\sim 0.1$ mag, while the earlier CASTLES observations 
show a larger shift and a significant wavelength dependence, suggesting that there 
is significant microlensing in this system. In any case further observations are needed to 
better understand the chromatic behavior in this system. 

\subsection {Q~0142$-$100}

The gravitational lens Q~0142$-$100 (UM 673) was discovered by 
Surdej et al. (1987). It is a doubly imaged quasar with components 
separated by $2\farcs2$. The source redshift is $z_S=2.72$ 
(MacAlpine $\&$ Feldman 1982), and absorption lines detected in the quasar spectra
suggest a lensing galaxy at $z_L=0.49$ (e.g., Surdej et al. 1987). 
Several authors have discussed the chromatic behavior
of this lens system (Falco et al. 1999; Wisotzki et al. 2004; 
El\'{i}asd\'ottir et al. 2006), but its nature is still a matter of discussion. 
The photometric data obtained by Nakos et al. (2005) possibly 
detected chromatic microlensing and renewed interest in this lens system. 
The narrowband observations of Q~0142$-$100 were obtained at the NOT on two nights separated by 
15 days (see Table \ref{log_NOT}). The best fit to the images was found using a de 
Vaucouleurs profile for the lensing galaxy, and point sources for the quasar images using 
the astrometry and structural models from Leh\'ar et al. (2000).

\begin{figure}
\begin{center}
\vspace{0.5 cm}
\includegraphics[scale=0.5]{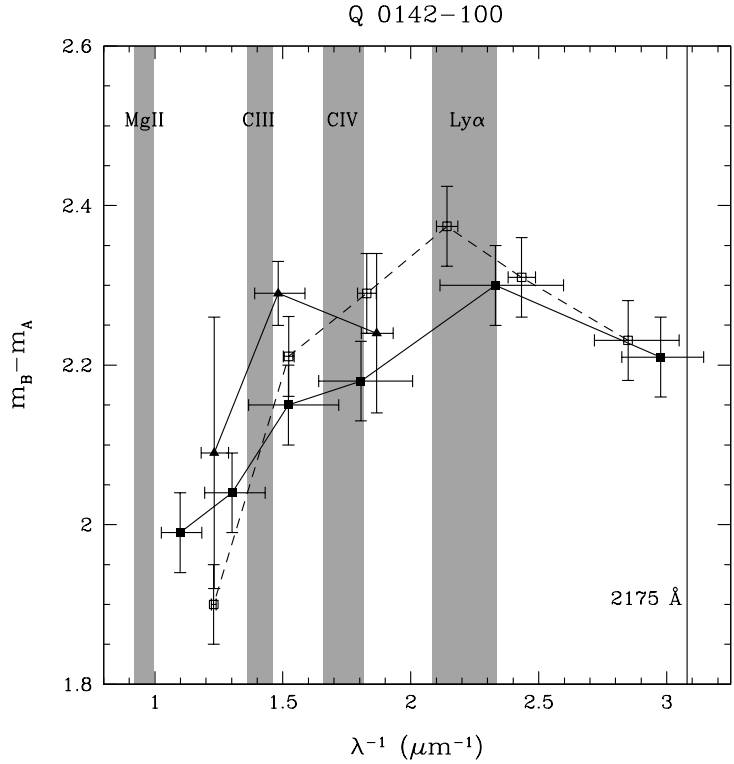}
\caption{\label{0142_phot} \small
Magnitude differences as a function of the inverse of the observed wavelength 
for Q~0142$-$100 at different epochs. The open squares correspond to the NOT data. The filled 
squares correspond to the data from El\'{i}asd\'{o}ttir et al. (2006), and the filled triangles are the data from CASTLES. 
The wavelengths corresponding to the most prominent quasar emission lines are indicated 
(from left to right: MgII, CIII], CIV and Ly$\alpha$). The horizontal error bars correspond 
to the filter widths. The solid vertical line corresponds to the expected 
position of the 2175 \AA ~feature.}
\end{center}
\end{figure}

Unfortunately, these observations shed little light on the chromatic behavior of Q~0142$-$100 
(open squares in Figure \ref{0142_phot}). Because most of the NOT filters are again contaminated 
in varying amounts by the emission lines of the quasar. The most
affected filters are the iac\#29 and Sty-b bands, which lie on the CIII] and Ly$\alpha$ 
emission lines. However, comparing our results with the data obtained 
by El\'{i}asd\'ottir et al. (2006) (filled squares in Figure \ref{0142_phot}) and by
CASTLES (Falco et al. 1999) (filled triangles), somewhat constrains the origin of the 
observed chromaticity. All three data sets, spanning a 10 year period, show the 
same color trends. Koptelova et al. (2010) also monitored this system during a 2 year 
period in $V$, $R$, and $I$ bands, and their average $m_B-m_A$ values for the different filters 
are consistent with our observations. This similarity between the color trends 
essentially eliminates the possibility that intrinsic source variability modulated by the time 
delay could produce the color trends, since the delay, while unknown, is small compared with the 
10 year interval between epochs. 

Extinction alone is also ruled out because the wavelengths of the observations 
correspond to the regime where all the extinction laws are roughly proportional to $\lambda$, with the 
2175 \AA~feature lying blueward of all the data, leaving nothing to create the observed 
parabolic wavelength dependence. Therefore the explanation for the observed chromaticity 
should be chromatic microlensing, even though the chromaticity changed little over 
a decade. Microlensing variations for Q~0142$-$100 are assured on the timescale $t_E\simeq 24$ 
years\footnote{This timescale was calculated using a concordance cosmology ($\Omega_m=0.3$, $\Omega_{\Lambda}=0.7$, 
$H_0=72$ km s$^{-1}$ Mpc$^{-1}$). The Einstein radius is $r_E \simeq 2.7 \times 10^{16}$ cm for 
0.3 $M_{\odot}$ microlenses, and the effective transverse velocity of the 
source (e.g., Kayser et al. 1986) is  $v\approx 360$ km s$^{-1}$ for a velocity dispersion 
of $\sigma_{\ast}=237$ km s$^{-1}$.} it should take the source to cross an Einstein radius, 
although fluctuations can occur on the shorter timescale of $t_S \simeq 3$ 
years it should take to cross the optical source size\footnote{We estimated the V-band disk size assuming 
that the disk emits like a thin disk (Shakura $\&$ Sunyaev 1973) and used the black hole 
mass of $M_{\mathrm{BH}}= 2.26 \times 10^{9} M_{\odot}$ estimated 
by Peng et al. (2006). For these assumptions the disk size in the V-band ($\lambda_{\mathrm{rest}}\simeq 0.15 \mu$m) 
is $R_V \simeq 2.9 \times 10^{15}$ cm.}. This suggests that the images must be lying in the broad, 
relatively structureless ``valleys'' between  the ``active ridges'' of the caustic networks. The redder 
color of image B may partly be due to contamination by the lens galaxy, since their separation is only $0.38$ arcsec 
and our data are limited by the resolution of our ground-based observations.

\section{Summary and Conclusions}

In this work, we studied the chromatic behavior of four lens systems 
using optical multi-wavelength data from a monitoring campaign performed 
at the NOT. The use of narrowband photometry under excellent seeing 
conditions has proved to be a powerful tool to disentangle 
chromatic microlensing from extinction, since depending on the source 
redshift, emission line contamination can be better separated from the 
continuum emission. 

In particular, the color variations in SDSS~J1650+4251 are probably dominated by extinction 
and require an extinction law without a 2175 \AA \ feature. This indicates that we 
have found a galaxy in which the extinction curve is similar to that of the SMC but at a cosmological redshift 
($z_{L}=0.58$). This result significantly increases the small sample of high-redshift galaxies with similar 
extinction features, and it is the first robustly determined from the pair method. 
This was possible due to the good wavelength resolution achieved by the use of narrow filters. 
The firm confirmation of an SMC-like dust at a cosmological distance is very important since many 
models including dust at higher redshifts assume the ``featureless'' SMC extinction law 
(e.g., Richards et al. 2003; Hopkins et al. 2004). 
The selection criteria would certainly be crucial in many fields 
of astrophysics, like, for instance, in the correct understanding of the expansion of the universe 
through SNe Ia (e.g., Jha et al. 2006, 2007).

We clearly detect chromatic microlensing in the image A of HE~0435$-$1223.  This single epoch observation, 
with microlensing $\Delta M_{R}=-0.19\pm0.04$ and chromatic microlensing $\Delta M_{b-R}=-0.13\pm0.08$, 
allows us to estimate the disk size and to constrain the power-law index in the size-wavelength scaling of 
the accretion disk. We found an R-band disk size of $r^{R}_s=13\pm5$ light days for a linear prior 
on $r^R_s$, and of $r^{R}_s=7\pm6$ light days for a logarithmic prior,  
and a value of $p=1.3\pm0.3$ consistent with the Shakura $\&$ Sunyaev (1973) thin disk model. Our result is in good 
agreement with those of other authors (Pooley et al. 2007, 2009; Morgan et al. 2010; Blackburne et al. 2011).

In the case of FBQ~0951+2635 and Q~0142$-$100, unfortunately, further observations 
are needed to completely understand the chromatic behavior of the systems, 
although our observations shed some light on their chromatic variations. The lack 
of chromaticity observed in the FBQ~0951+2635 NOT data is compatible with the 
absence of extinction, and the different chromatic behaviors observed 
at different epochs suggest that there is significant microlensing in the system. In the 
case of Q~0142$-$100, microlensing scenarios are also favored to explain the observed 
chromaticity, although wavelength-dependent contamination by the lens galaxy in the flux of image B, due to 
its proximity, is not ruled out.

\bigskip

\noindent Acknowledgments:
 
This research was supported by the European Community's Sixth Framework Marie Curie 
Research Training Network Programme, Contract No. MRTN-CT-2004-505183 ``ANGLES'', and by 
the Spanish Ministerio de Educaci\'{o}n y Ciencias (grants AYA2004-08243-C03-01/03  
and AYA2007-67342-C03-01/03). A.M.M. acknowledges the support of Generalitat Valenciana, 
grant APOSTD/2010/030. J.A.M. is also supported by the Generalitat Valenciana with the 
grant PROMETEO/2009/64. C.S.K. is supported by NSF grants AST-0708082 and AST-1009756.


\clearpage

\clearpage

\clearpage

\clearpage

\clearpage

\clearpage

\clearpage

\clearpage

\clearpage

\clearpage

\clearpage

\clearpage

\clearpage

\clearpage

\clearpage

\clearpage

\begin{deluxetable}{lcrc}
\tabletypesize{\scriptsize}
\tablecaption{\scriptsize \label {log_NOT}log of ALSFOC Observations}
\tablewidth{0pt}
\tablehead{
\colhead{Target\tablenotemark{~}} & \colhead{Observation Date} & \colhead{Filter} & \colhead{Exposure (s)}}
\startdata
SDSS J1650+4251 & 2003 Aug 26 & Str-u ($\lambda=3510$ \AA)& $4\times300$ \\
SDSS J1650+4251 & 2003 Aug 26 & I-band ($\lambda=8130$ \AA) & $3\times900$ \\
SDSS J1650+4251& 2003 Aug 28 & Str-b ($\lambda=4670$ \AA) & $3\times600$ \\
SDSS J1650+4251& 2003 Aug 28 & Str-y ($\lambda=5470$ \AA) & $3\times600$ \\
SDSS J1650+4251& 2003 Aug 28 & Iac\#28 ($\lambda=6062$ \AA) & $3\times600$ \\
SDSS J1650+4251& 2003 Aug 28 & H$\alpha$ ($\lambda=6567$ \AA) & $3\times600$ \\
SDSS J1650+4251& 2003 Aug 28 & Iac\#29 ($\lambda=7015$ \AA) & $3\times600$\\
SDSS J1650+4251& 2003 Aug 28 & I-band ($\lambda=8130$ \AA) & $1\times300$ \\
HE~0435$-$1223   & 2007 Oct 15 & Str-v ($\lambda=4110$ \AA) & $3\times600$ \\
HE~0435$-$1223   & 2007 Oct 15 & Str-y ($\lambda=5470$ \AA) & $3 \times600$\\
HE~0435$-$1223   & 2007 Oct 15 & H$\alpha$ ($\lambda=6567$ \AA) & $3\times600$ \\
HE~0435$-$1223   & 2007 Oct 15 & I-band ($\lambda=8130$ \AA) &  $4\times100$\\
HE~0435$-$1223   & 2007 Oct 31& Str-u ($\lambda=3510$ \AA)& $4\times600 $\\
HE~0435$-$1223   & 2007 Oct 31& Str-v ($\lambda=4110$ \AA) & $3\times600$ \\
HE~0435$-$1223   & 2007 Oct 31 & Str-b ($\lambda=4670$ \AA) & $3\times600$ \\
HE~0435$-$1223   & 2007 Oct 31 & Str-y ($\lambda=5470$ \AA) & $3 \times600$\\
HE~0435$-$1223   & 2007 Oct 31 & Iac\#28 ($\lambda=6062$ \AA) & $3\times600$ \\
HE~0435$-$1223   & 2007 Oct 31& H$\alpha$ ($\lambda=6567$ \AA) & $3\times600$ \\
HE~0435$-$1223   & 2007 Oct 31 & Iac\#29 ($\lambda=7015$ \AA) & $3\times600$\\
HE~0435$-$1223   & 2007 Oct 31 & I-band ($\lambda=8130$ \AA) &  $4\times100$\\
FBQ~0951+2635  & 2006 Nov 17 & Str-u ($\lambda=3510$ \AA)& $4\times600 $\\
FBQ~0951+2635  & 2006 Nov 17 & Str-v ($\lambda=4110$ \AA) & $3\times600$ \\
FBQ~0951+2635  & 2006 Nov 17 & Str-b ($\lambda=4670$ \AA) & $3\times600$ \\
FBQ~0951+2635  & 2006 Nov 17 & Str-y ($\lambda=5470$ \AA) & $3 \times600$\\
FBQ~0951+2635  & 2006 Nov 17 & Iac\#28 ($\lambda=6062$ \AA) & $3\times600$ \\
FBQ~0951+2635  & 2006 Nov 17 & Iac\#29 ($\lambda=7015$ \AA) & $3\times600$\\
FBQ~0951+2635  & 2006 Nov 17 & I-band ($\lambda=8130$ \AA) &  $4\times100$\\
Q~0142$-$100     & 2006 Sep 23 & Str-u ($\lambda=3510$ \AA)& $4\times600 $\\
Q~0142$-$100     & 2006 Sep 23 & Str-v ($\lambda=4110$ \AA) & $3\times600$ \\
Q~0142$-$100     & 2006 Sep 23 & Str-b ($\lambda=4670$ \AA) & $3\times600$ \\
Q~0142$-$100     & 2006 Sep 23 & Str-y ($\lambda=5470$ \AA) & $3 \times600$\\
Q~0142$-$100     & 2006 Sep 23 & H$\alpha$ ($\lambda=6567$ \AA) & $3\times600$ \\
Q~0142$-$100     & 2006 Sep 23 & I-band ($\lambda=8130$ \AA) &  $4\times100$\\
Q~0142$-$100     & 2006 Nov 17 & Str-u ($\lambda=3510$ \AA)& $4\times600 $\\
Q~0142$-$100     & 2006 Nov 17 & Str-b ($\lambda=4670$ \AA) & $3\times600$ \\
Q~0142$-$100     & 2006 Nov 17 & Iac\#28 ($\lambda=6062$ \AA) & $3\times600$ \\
Q~0142$-$100     & 2006 Nov 17 & Iac\#29 ($\lambda=7015$ \AA) & $3\times600$\\
Q~0142$-$100     & 2006 Nov 17 & I-band ($\lambda=8130$ \AA) &  $4\times100$
\enddata
\tablenotetext{~}{Str is short for Str\"omgren.}
\tablenotetext{~}{Iac\#28 and Iac\#29 are non-standard filters shared with the IAC-80 telescope 
(Observatorio del Teide, Tenerife, Spain). These two filters have an FWHM of 20 nm and 9.5 nm, respectively.}
\end{deluxetable}

\begin{deluxetable}{lccc}
\tabletypesize{\small}
\tablecaption{\label {log_LT} log of RATCam Observations}
\tablewidth{0pt}
\tablehead{
\colhead{Target} & \colhead{Observation Date} & \colhead{Filter} & \colhead{Exposure (s)}}
\startdata
SDSS J1650+4251 & 2006 Jun 14 & u' ($\lambda=3500$ \AA) &  $7\times120$ \\
SDSS J1650+4251 & 2006 Jun 14 & g' ($\lambda=4800$ \AA) & $4\times100$ \\
SDSS J1650+4251 & 2006 Jun 14 & r' ($\lambda=6250$ \AA) & $2\times100$ \\
SDSS J1650+4251 & 2006 Jun 14 & i' ($\lambda=7700$ \AA) & $2\times100$ \\
SDSS J1650+4251 & 2006 Jun 14 & z' ($\lambda=9100$ \AA) & $3\times120$ \\
SDSS J1650+4251 & 2006 Jun 24 & u' ($\lambda=3500$ \AA)&  $3\times120$ \\
SDSS J1650+4251 & 2006 Jun 24 & g' ($\lambda=4800$ \AA) & $3\times100$ \\
SDSS J1650+4251 & 2006 Jun 24 & r' ($\lambda=6250$ \AA) & $2\times100$ \\
SDSS J1650+4251 & 2006 Jun 24 & i' ($\lambda=7700$ \AA) & $1\times100$ \\
SDSS J1650+4251 & 2006 Jun 24 & z' ($\lambda=9100$ \AA) & $3\times120$ \\
SDSS J1650+4251 & 2006 Jun 30 & u' ($\lambda=3500$ \AA)&  $1\times120$ \\
SDSS J1650+4251 & 2006 Jun 30 & g' ($\lambda=4800$ \AA) & $2\times100$ \\
SDSS J1650+4251 & 2006 Jun 30 & r' ($\lambda=6250$ \AA) & $1\times100$ \\
SDSS J1650+4251 & 2006 Jun 30 & i' ($\lambda=7700$ \AA) & $5\times100$ \\
SDSS J1650+4251 & 2006 Jul 7 & u' ($\lambda=3500$ \AA)&  $3\times120$ \\
SDSS J1650+4251 & 2006 Jul 7 & g' ($\lambda=4800$ \AA) & $5\times100$ \\
SDSS J1650+4251 & 2006 Jul 7 & r' ($\lambda=6250$ \AA) & $4\times100$ \\
SDSS J1650+4251 & 2006 Jul 7 & i' ($\lambda=7700$ \AA) & $4\times100$ \\
SDSS J1650+4251 & 2006 Jul 7 & z' ($\lambda=9100$ \AA) & $4\times120$ \\
\enddata
\end{deluxetable}

\begin{deluxetable}{rc}
\tabletypesize{\small}
\tablecaption{\small \label {LT_phot_1650} SDSS~1650+4251 LT PHOTOMETRY}
\tablewidth{0pt}
\tablehead{
\colhead{Filter} & \colhead{$m_B-m_A$}}
\startdata
u'($\lambda=3500$ \AA) & 2.12$\pm$0.05 \\
g'($\lambda=4800$ \AA) & 1.90$\pm$0.05\\
r'($\lambda=6250$ \AA) & 1.76$\pm$0.05\\
i'($\lambda=7700$ \AA) & 1.70$\pm$0.05\\
z'($\lambda=9100$ \AA) & 1.59$\pm$0.06\\
\enddata
\end{deluxetable}

\def\foo{\hphantom{-}}
\begin{deluxetable}{lccccccccc}
\tabletypesize{\scriptsize}
\rotate
\tablecaption{\scriptsize \label {NOT_phot_1650} NOT-PHOTOMETRY}
\tablewidth{0pt}
\tablehead{
\colhead{Target} & \colhead{$\Delta m$} & \colhead{Str-u} &\colhead{Str-v} &\colhead{Str-b} & \colhead{Str-y} & \colhead{Iac\#28} 
& \colhead{H$\alpha$} & \colhead{Iac\#29} &\colhead{I-band}}
\startdata
SDSS~1650+4251 & $m_B-m_A$ & $\foo2.37\pm0.09$  & $\foo2.25\pm0.09$ & $\foo2.12\pm0.09$ & $\foo2.00\pm0.09$ & $\foo1.95\pm0.09$ & $\foo1.90\pm0.09$ & 
$\foo1.85\pm0.09$ & $\foo1.93\pm0.09$\\
HE~0435$-$1223\tablenotemark{*}   & $m_B-m_A$&        &   $\foo0.65\pm0.02$  &     &  $\foo0.71\pm0.01$ &      &   $\foo0.66\pm0.02$  &   &  $\foo0.61\pm0.05$\\
               & $m_B-m_C$&     &  $-0.01\pm0.03$  &     & $\foo0.01\pm0.01$ &      &   $\foo0.04\pm0.03$  &   &  $\foo0.07\pm0.05$ \\
               & $m_B-m_D$&     &  $\foo0.15\pm0.03$  &     &  $\foo0.16\pm0.01$ &      &   $\foo0.22\pm0.02$  &   &  $\foo0.22\pm0.04$ \\
HE~0435$-$1223\tablenotemark{**}  & $m_B-m_A$ & $\foo0.87\pm0.14$ & $\foo0.69\pm0.02$ & $\foo0.81\pm0.08$ & $\foo0.72\pm0.09$ & $\foo0.72\pm0.10$ & 
$\foo0.68\pm0.02$  & $\foo0.67\pm0.07$ & $\foo0.61\pm 0.04$\\
               & $m_B-m_C$     & $\foo0.06\pm 0.14$ & $-0.03\pm0.03$ & $\foo0.03\pm0.09$ & $\foo0.01\pm0.11$ & $\foo0.05\pm0.09$&
$\foo0.06\pm0.02$  & $\foo0.04\pm0.07$ & $\foo0.07\pm0.03$\\
               & $m_B-m_D$&   $-0.06\pm0.13$ & $-0.24\pm0.02$ & $-0.13\pm0.06$ & $-0.21\pm0.07$& $-0.20\pm0.07$ 
& $-0.23\pm0.02$ & $-0.24\pm0.06$ & $-0.19\pm0.05$\\
FBQ~0951+2635  & $m_B-m_A$ &  $\foo1.30\pm0.05$& $\foo1.42\pm0.05$ & $\foo1.37\pm0.05$ & $\foo1.44\pm0.05$ & $\foo1.33\pm0.05$ &   & $\foo1.36\pm0.05$ & $\foo1.43\pm0.05$\\
Q~0142$-$100\tablenotemark{\dag}     & $m_B-m_A$& $\foo2.23\pm0.05$ & $\foo2.31\pm0.05$ &   $\foo2.37\pm0.05$  & $\foo2.29\pm0.05$  &             &      $\foo2.21\pm0.05$    &   &  $\foo2.39\pm0.05$ \\
Q~0142$-$100 \tablenotemark{\dag \dag}  & $m_B-m_A$& $\foo2.24\pm0.05$ &     &  $\foo2.36\pm0.05$ &    & $\foo2.18\pm0.05$ &    &  $\foo2.37\pm0.05$  & $\foo2.36\pm0.05$ \\

\enddata

\tablenotetext{*}{Observation date: 2007 October 15.}
\tablenotetext{**}{Observation date: 2007 October 31.}
\tablenotetext{\dag}{Observation date: 2006 September 23.}
\tablenotetext{\dag \dag}{Observation date: 2006 November 17.}
\end{deluxetable}

\def\foo{\hphantom{-}}
\begin{deluxetable}{lcccccc}
\tabletypesize{\footnotesize}
\tablecaption{\label {pos_1650}SDSS~J1650+4251 Component Positions.}
\tablewidth{0pt}
\tablehead{\colhead{COMPONENT} &
\multicolumn{2}{c}{WIYN\tablenotemark{*}}&
\multicolumn{2}{c}{NOT}\\
&$\Delta$R.A.&$\Delta$Decl.&
$\Delta$R.A.&$\Delta$Decl.\\
}
\startdata
Image A &$\foo\equiv0$ &$\foo\foo\equiv0$ &$\foo\equiv0$&$\foo\foo\equiv0$\\
Image B &$0.223\pm0.002$&$\foo 1.163\pm0.001$&$0.24\pm0.02$&$\foo 1.17\pm0.02$\\
Lens galaxy G&$0.017\pm0.032$&$-$$0.872\pm0.026$&$0.015\pm0.001$&$-$$0.88\pm0.07$\\
\enddata

\tablenotetext{*}{Relative positions to image A obtained by Morgan et al. (2003)
with the WIYN 3.5m telescope at the Kitt Peak National Observatory.}

\end{deluxetable}

\def\foo{\hphantom{7}}
\begin{deluxetable}{rccccccc}
\tabletypesize{\small}
\tablecaption{\small \label {tab_Rpa} HE~0435$-$1223 R-band Disk Size and Power-law Index Estimations 
for $\alpha<0.15$}
\tablewidth{0pt}
\tablehead{
\colhead{$~$}& \multicolumn{2}{c}{Linear Prior}&
\multicolumn{2}{c}{Logarithmic Prior}\\
$\alpha\foo$&\foo$r^R_s$ (light days) & \foo$p$ &
\foo$r^R_s$(light days)  & \foo$p$ &\\
}
\startdata
0.01 & $\foo9\pm2$ & $1.3\pm0.3$ &$9\pm2$  & $\foo\gtrsim 2$\\  
0.05 &$ 13\pm4$ & $1.3\pm0.3$ & $7\pm6$ &$1.7\pm0.3$\\ 
0.1  & $\foo7\pm6$ &  $1.3\pm0.3$ & $7\pm4$ &$1.3\pm0.3$\\
0.15 & $11\pm6$ &  $1.7\pm0.3$ & $9\pm4$ &$1.7\pm0.3$
\enddata
\end{deluxetable}

\end{document}